\documentclass[apj]{emulateapj}
\usepackage{apjfonts}

\newcommand{\lya}{Ly$\alpha$}
\newcommand{\sfr}{M$_{\odot}$yr$^{-1}$}
\newcommand{\sfrd}{M$_{\odot}$yr$^{-1}$Mpc$^{-3}$}

\shorttitle{Submm observatons  of LAEs}
\shortauthors{Webb et al.}

\begin{document}


\title{Deep Submillimeter Observations of Two Lyman-$\alpha$ Emitting Galaxies at $z\sim$ 6.5}


\author{T.M.A. Webb\altaffilmark{1}}
\email{webb@physics.mcgill.ca}

\altaffiltext{1}{Department of Physics, McGill University, Rutherford Physics Building, 3600 rue University, Montr\'eal, Canada, H3A 2T8} 

\author{K.-V.H. Tran\altaffilmark{2,3,4}}
\author{S.J. Lilly\altaffilmark{4}}
\altaffiltext{2}{NSF Astronomy \& Astrophysics Postdoctoral Fellow}
\altaffiltext{3}{Harvard-Smithsonian Center for Astrophysics, 60
  Garden Street, Cambridge, MA 02138}
\altaffiltext{4}{Institute for Astronomy, ETH Z\"urich, CH-8093
  Z\"urich, Switzerland} 
\author{P. van der Werf\altaffilmark{1}}







\begin{abstract}

We present deep submillimeter imaging of two spectroscopically
confirmed $z\sim$ 6.5 Lyman-$\alpha$ emitters (LAEs) in the Subaru
Deep Field.  Although we reach the nominal confusion limit at
850{\micron}, neither LAE is detected at 850{\micron} nor
450{\micron}, thus we conclude that the LAEs do not contain large dust
masses ($<$ 2.3 $\times$ 10$^8$ M$_\odot$ and  $<$ 5.7 $\times$ 10$^8$ M$_\odot$).  The limit on their average
L$_\mathrm{FIR}$/L$_\mathrm{UV}$ ratios ($\lesssim$ 35) is
substantially lower than seen for most submillimeter selected galaxies
at $z\sim$ 3, and is within the range of values exhibited by
Lyman-break galaxies.  We place upper-limits on their individual star
formation rates of $\lesssim$ 248 M$_{\odot}$yr$^{-1}$ and $\lesssim$ 613 M$_{\odot}$yr$^{-1}$, and on the
cosmic star formation rate density of the $z\sim$ 6.5 LAE
population of $\lesssim$ 5.0$\times$10$^{-2}$ {\sfrd}.  In the two
submm pointings, we also serendipitously detect seven sources at
850$\mu$m that we estimate to lie at $1<z<5$.

\end{abstract}





\section{Introduction}

Deep narrow-band imaging at optical wavelengths has identified a
sizable number of star-forming galaxies at $z\sim$ 6.5 through searches
for strong Lyman-$\alpha$ (Ly$\alpha$) emission
\citep{aji03,kod03,hu02,rho04,tan05}.  At this redshift, the universe
was less than 10$^9$ years old and these early systems provide insight
into many important aspects of the formation of galaxies at very early
times.  Although the sample size of $z>$ 6 galaxies is increasing
through the efforts of large surveys, most efforts are focused on
optical and near-infrared observations and follow-up studies at longer
wavelengths remain scarce.

The detection of $z\sim$ 6.5 star-forming galaxies at submm or
far-infrared wavelengths would have a number of important
implications.  Firstly, detection at these wavelengths requires that
these young systems contain substantial amounts of dust and, at $<$
10$^9$ years after the Big Bang, would provide 
current dust production models with strong constraints.  Although large dust masses have been
found in high redshift ($z\gtrsim4$) quasars and radio galaxies
\citep[e.g.][]{omo01, prid01,arc01,rob04}, 
quasar winds offer an explanation for the production of their dust \citep{elv04}; however, such a mechanism cannot be
invoked for pure starburst galaxies, $i.e.$ the $z\sim6.5$ Lyman$\alpha$ emitters (LAEs).
Because the age of the universe at $z\sim6.5$ is comparable to the
evolutionary timescales of low-mass stars, evolving stars cannot be
the primary source of dust.  Rather, these star-forming galaxies can
only manufacture dust via grain formation in Type II supernovae (SNe)
\citep{tod01,mor03}; even so, a top-heavy Initial Mass Function and
very high star formation efficiency also must be invoked.

Secondly, quantifying the amount of dust in galaxies at $z\sim$ 6.5 is
important for measuring their individual star formation rates,
bolometric luminosities, and by extension, the cosmic star formation
rate density (SFRD) at $z>$ 6.  The SFRD has now been measured for
large samples of galaxies to $z\sim$ 5 and thus far does not exhibit
any clear sign of turning over \citep{lil95,mad96,con97,ste99,gia05};
however, this must occur somewhere in the relatively short amount of
time between $z\sim$ 5 and the epoch of reionization. Currently the
majority of measurements at $z\gtrsim$ 6 rely on a relatively small
number of emission line galaxies found through narrow-band optical
searches for Ly$\alpha$ emission \citep{hu02,rho04,tan05}.  While
these studies measure a SFRD that is apparently an order of magnitude
less than at $z\sim$ 5, their accuracy is hindered by small numbers,
uncertainties in the completeness corrections, and the unknown amount
of dust extinction at $z > 3$. 

The unknown correction for dust at $z > $ 6 is potentially substantial, as
illustrated by recent submm surveys
\citep[e.g.,][]{sma97,lil99,web03b,cha05} that show the correction due
to dust is large at $z\sim$ 3: the dust-uncorrected optical/UV
estimates of the SFR are one to two orders of magnitude lower than the
infrared estimates, and the total star formation in rare,
infrared-selected objects is comparable to that seen in large
populations of optically-selected galaxies \citep{cha05}. Thus, if
there exists a population of heavily dust-obscured star-forming
galaxies at $z\sim$ 6.5, infrared observations are crucial for
measuring their total individual and integrated star formation rates.
 
One might assume that because dust is an efficient absorber of
Ly$\alpha$ photons, the existence of strong Ly$\alpha$ emission in
these systems precludes the presence of significant amounts of dust.
However, this does not empirically appear to be the case: thus far,
$\sim30$\% of the spectroscopic redshifts for submm-detected galaxies
are, in fact, derived from strong {\lya} in emission \citep{cha05}.
Only a small number of LAEs at
$z\sim$ 2 have been observed in the submm, and though none of these
have been detected \citep{bar99,sma03} we note that
many of the larger, extended Lyman-$\alpha$ blobs (LABs) have been
detected in the submm and/or infrared \citep{cha01,gea05,col06}.

Here we extend for the first time the search for submm emission to two
LAEs at $z$ = 6.5.  We are aided by the negative $K$-correction at
submm wavelengths that results in equal sensitivity for detecting
dust-obscurred star formation at $z\sim6$ as at $z\sim1$.  The two
targeted LAEs were discovered through narrow band imaging at
${\lambda_c}$=9196{\AA} in the Subaru Deep Field
\citep[SDF;][]{kod03}, and have been spectroscopically confirmed to
lie at $z=$ 6.541 (SDF-1) and 6.578 (SDF-2).  We present results at
850 and 450{\micron} for both LAEs; while neither object is detected,
we attempt to place upper limits on the properties of the $z\sim6.5$
LAE population using our deep submm observations.  The paper is
organized as follows: in \S2 we describe the observations and data
analysis; \S3 presents the derived properties of the $z\sim6.5$ LAEs;
and \S4 a brief analysis of the serendipitous detections.  Our
conclusions are in \S5. We assume an ${\Omega}_\mathrm{M}$=0.3,
${\Omega}_{\Lambda}$=0.7 cosmology with H$_{\circ}$=70 km/s/Mpc
throughout.

\section{Data and Analysis}
 
 \subsection{Observations and Data Reduction}

We observed the objects using the Submillimeter (submm) Common-User
Bolometer Array \citep[SCUBA;][]{hol99} on the James Clerk Maxwell
Telescope (JCMT) over eight partial nights in 2003, 2004, 2005.  We
obtained data at 850{\micron} and 450{\micron} simultaneously, but due
to beam instabilities and poor sky transmission the 450{\micron} data
are of poorer quality than the 850{\micron} data.  To fill in the
under-sampled sky, SCUBA was stepped through a regular 64-point jiggle
pattern and sky flux was removed to first order by employing a 3-point
chop with a chop-throw of 30{\arcsec}. The chop position angle was
held constant in right ascension which results in a characteristic
negative-positive-negative beam shape on the final map for real point
sources.  Telescope pointing was checked every 1.5 hours, and sky
opacity was monitored through sky dips every 1.5-3.0 hours and in real
time along the line of sight using the JCMT water vapor meter when
operational.

The data were reduced using a combination of general SURF
\citep{jen99} routines and custom IDL programs written by ourselves.
We first applied standard flat-fielding and extinction corrections.
Although the 3-point chop employed by SCUBA removed sky to first
order, residual sky flux remained; SURF attempts to remove this
through a subtraction of a single median or average sky measurement
for each second of measurement.  To improve on this, we subtracted a
sky plane fit to each second of the data time streams and all
bolometers, with the bolometers weighted by their individual noise.
In a single second of integration, any structure in the maps is only
due to sky flux and thus this does not add any systematic offset.  A
comparison of images produced with and without this method indicates a
decrease in the noise of $\sim$30\%. Noise spikes were then
iteratively removed using a simple standard deviation clipping routine
(at 3$\sigma$ using the entire array) and the data were regridded onto
an astrometrically calibrated map using standard Gaussian weighting.

Total integration times (including the off-source chop time) were 47ks
for SDF-1 and 19ks for SDF-2.  The final images have central
(unsmoothed) noise values of rms=1.2mJy/beam (SDF-1) and
rms=1.6mJy/beam (SDF-2); thus both pointings reach depths close to the nominal confusion
limit at 850$\micron$. Note that the noise does not scale perfectly as
$t^{-1/2}$ due to weather differences and night to night variations in the noise properties of the bolometer array.  The jiggle-pattern and
rotation of the Nasmyth-located SCUBA on the sky results in uneven
coverage of the sky, with the center of the image receiving more
integration time than the outer edges and hence the centers of each
image, where the LAEs are located, are the deepest points on each map.

\subsection{Source Detection}

To improve point source recovery, we have employed the source
detection and extraction technique of \citet{eal00} and
\citet{web03b}.  In short, the map is iteratively cleaned using a beam
template generated from the point source calibration maps.  Sources
are detected by convolving the cleaned map with the beam
template. Such a technique takes advantage of the 
negative-positive-negative pattern that real sources exhibit, but that
noise spikes do not, and improves the separation of blended sources.

The noise as a function of position in each map was determined in the
following way.  We produced fake bolometer time-streams, drawn from
the real data using a bootstrapping technique.  These were passed
through our analysis pipeline to produce 1000 full map realizations
and the noise is taken to be the rms variation of each pixel in these
maps.  This technique agrees well with Monte Carlo simulations of the
noise, assuming Gaussian statistics \citep{web03b,web05}.

\subsubsection{SDF-1 \& SDF-2}

\begin{figure}
\plotone{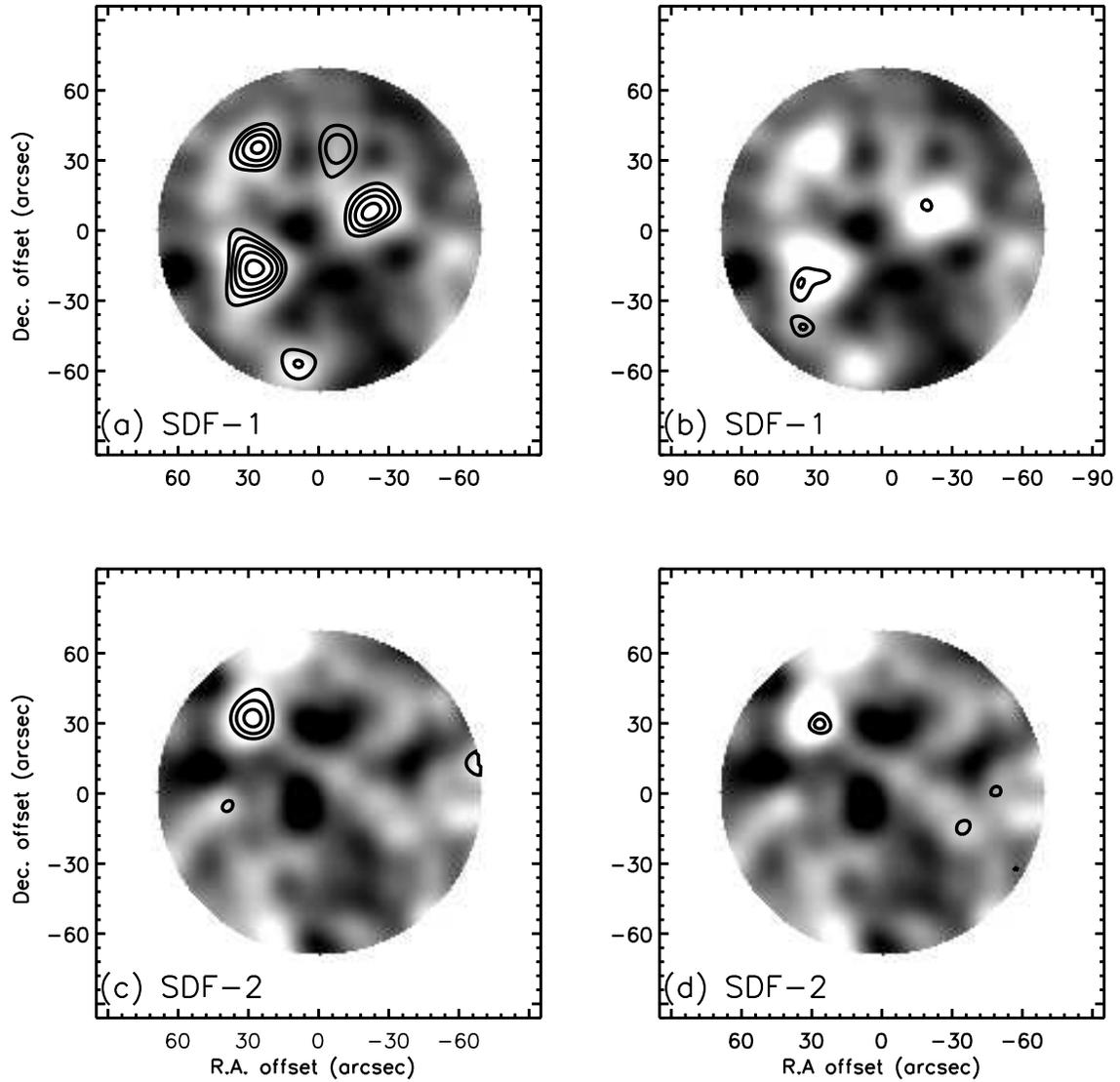}
\caption{The submm maps of the two LAE fields. The grey scale corresponds
to flux density, with white denoting positive flux,  smoothed with a 14{\arcsec} Gaussian.  All four images
correspond to 850$\mu$m; the top two panels show SDF-1 and the bottom
two show SDF-2.  Overlaid on the left two panels (a,c) are the
850$\mu$m S/N contours, determined through our iterative cleaning
algorithm. The right two panels show 450$\mu$m contours overlaid on
the 850$\mu$m maps to illustrate positional coincidence between the
two wavelengths.  In all four panels, the contour levels start at
3$\sigma$ and increase in 1$\sigma$ steps.  East is to the left and north is to the top.  \label{submm}}
\end{figure}

\begin{figure}
\plotone{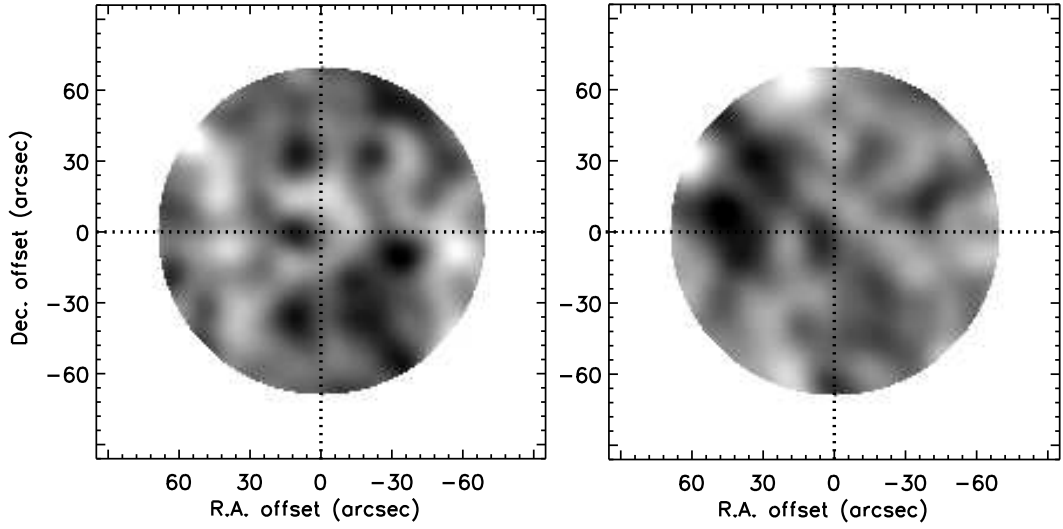}
\caption{The submm images centered on each LAE, after the iterative
cleaning has removed all $>$3.5$\sigma$ sources: the left panel
corresponds to SDF-1 and the right panel to SDF-2.  The LAEs lie at
(0,0) in both images (denoted by the intersection of the dotted
lines), and neither are detected at 850{\micron} or 450{\micron}.  The
measured flux at (0,0) in each image is listed in
Table~\ref{laetab}. East is to the left and north is to the top. \label{submme}}
\end{figure}

In Fig.~\ref{submm}, we show the two LAE fields at 850{\micron} after
smoothing with a 14{\arcsec} Gaussian; the LAEs are at the center of
each image. Overlaid are the signal-to-noise ratio (S/N) contours
resulting from the iterative cleaning procedure for the 850{\micron}
and 450{\micron} maps; the 450{\micron} contours are shown over the
850{\micron} grey-scale map to aid in positional comparison.  Neither
image contains significant flux at the locations of the LAEs.
Fig.~\ref{submme} shows the two 850{\micron} maps after the
$>$3.5$\sigma$ sources have been cleaned and removed, with the LAE
position marked and the flux at this position listed in
Table~\ref{laetab}.

Although the maps have a mean flux of zero, there is residual
 positive and negative structure which is not only due to noise but also 
 real structure in the extragalactic submm sky below the
confusion limit of SCUBA. This combination of noise and source confusion
results in an offset between the true and recovered positions of
objects; simulations show that sources at $\sim$3.5-4$\sigma$ are
rarely recovered at their original location and can be shifted up to 8{\arcsec} away \citep[$\sim$ 90-95 percentiles][]{web03b,ivi05}.  Thus,
if either of the LAEs exhibited significant submm flux, one would
expect it to be offset from the known position by a few arcseconds;
however, there are no significant sources within any reasonable search
radius of the LAEs.

To determine robust upper limits on the observed flux of the LAEs, we
placed and recovered sources at the LAE position in each map,
following the same cleaning procedure described above.  Fake sources
were deemed recovered if they were detected at $>$3$\sigma$ in the
cleaned map within 8{\arcsec} of the original input position.  For the
SDF-1 field, this analysis indicated a detection threshold that was in
good agreement with the limit expected from the noise map. Although
the LAE position is coincident with the off-beam of a bright (3.8mJy)
source, we were able to separate the contributions from this and the
input (fake) source and recover the fake source at $\geq$ 3$\sigma$
within 8{\arcsec} of the input position, provided the input flux was
$>$1.5mJy.

The SDF-2 map is shallower than SDF-1 and appears to suffer from
greater source confusion in its central region. Although the LAE is
not confused with a bright neighbor, structure beneath the 3$\sigma$ detection
limit  makes it
difficult to recover flux at the LAE position. Specifically, it lies in a negative region 
of the map (which is not statistically significant), and although this 
minimum could be the off-beam of one or more faint
neighbors, they lie below the confusion limit and thus cannot be cleaned from the map.
 Our simulations
indicate that we are unable to recover sources at this position above 3$\sigma$ below
an input flux of 3.8mJy to within 8{\arcsec} of the input position. 
These recovery limits are listed in Table~\ref{laetab} for 850$\mu$m and
450$\mu$m where both were determined in the manner described above.

\subsubsection{Serendipitous Detections}

Although the $z\sim6.5$ LAEs are not detected at submm wavelengths,
the deep pointings reveal multiple sources near the LAEs; the sources
detected in the two fields are listed in Table~\ref{submmtab}.  SDF-1
contains five sources detected above 3.5$\sigma$ at 850$\mu$m while
SDF-2 contains two such sources.  Of these seven 850$\mu$m sources,
two are also cleanly detected at 450$\mu$m with S/N$>$ 3.5 and
positional offsets of 4.0$\arcsec$.  A third 850$\mu$m source has two
possible 450$\mu$m counterparts: a 4.0$\sigma$ detection 10$\arcsec$
from the 850$\mu$m position, or a 2.8$\sigma$ peak offset by
5.3$\arcsec$.  Because a 10{\arcsec} offset between the 450$\micron$ and 850{\micron} positions
is statistically unlikely,   we assume that
the closer 2.8$\sigma$ peak corresponds to the 450$\mu$m emission for this
source; however, we note that none of our final conclusions are
greatly altered if the more distant object was the correct
counterpart. Also listed in Table~\ref{submmtab} are three additional
sources detected only at 450$\mu$m (all at $>$3.5$\sigma$).

\section{Lyman$-\alpha$ Emitting Galaxies at $z\sim6.5$}

\subsection{Star Formation Rates}

A galaxy's far-infrared luminosity is due to thermally radiating dust
and is correlated with the galaxy's current rate of star formation,
provided the energy contribution from an active galactic nucleus (AGN)
is negligible \citep{ken98,bel03}. Locally, the most intense starburst
galaxies are also the most luminous in the infrared; however, these
systems do not show a simple relation between infrared excess and UV
slope \citep[the IRX-$\beta¬$ relation;][]{gol02}.  Therefore
extrapolating from the UV to a total L$_\mathrm{bol}$ and
corresponding star formation rate is not trivial, and in many cases it
may not even be possible.

This situation is mirrored at high redshift ($z\sim$ 2-3) where
luminous (L $>$ 10$^{11}$ L$_\odot$), dusty starbursts are more
prevalent \citep[e.g.,][]{bor02,web03b}. For such systems, the rest-frame UV
estimated SFR can be two orders of magnitude less than that implied
by the rest-frame far-infrared emission \citep{cha05}.  Far-infrared imaging
offers an independent and, for the most luminous systems, a more
accurate measure of their total star formation rates and bolometric
luminosities.  Moreover, because of the flat flux-redshift relation
for infrared luminous galaxies at 850$\mu$m beyond $z\sim$ 0.5, such
systems are equally detectable at $z\sim$ 6 as they are at $z\sim$ 1,
unlike  optical and near-infrared observations.

However, converting infrared luminosities to star formation rates
suffers from a number of uncertainties.  While the empirically
measured local SFR-L$_{\mathrm{IR}}$ relation does hold over four
orders of magnitude in L$_\mathrm{IR}$, the scatter about this
relation is approximately a factor of two.  At high redshifts, the
scatter could be much greater and/or systematically offset from the
local relation as uncertainties due to differences in the Initial Mass
Function \citep[IMF;][]{bau05}, differences in dust properties such as
temperature or emissivity, and the contamination from AGN increase.
This is especially relevant at the redshifts considered here where we
are beginning to probe the very first generation of stars and dust.

With these caveats in mind, we have employed the following method to
convert our limit on the observed $S_\mathrm{850{\mu}m}$ to a star
formation rate.  We adopt a modified black-body spectral energy
distribution (SED) with a temperature of 36K, a dust emissivity of
$\beta$=1.5.  Following \citet{bla03} we apply a Wien correction
short-ward of 80$\mu$m with a slope of $\alpha$=2.9. These parameters
are chosen to match the measurements of the submillimeter luminous
population at $z\sim$ 3 \citep{cha05}, but they are also in good
agreement with the results from local ULIRG galaxies
\citep{dun00,bla03}. Using this SED, we calculate L$_\mathrm{IR}$ (i.e., L(8-1000$\mu$m))
for the two LAEs and use the relation from \citet{bel03} to convert
this to a star formation rate.  The results are included in
Table~\ref{submmtab}.  Differences in the assumed
temperature of $\sim$30\% (the inter-quartile range of Chapman et al.)
result  in a difference in the inferred star
formation rates of $\sim$70\%, in the sense that assuming hotter dust results in
larger a infrared luminosity and therefore a larger inferred star formation rate.

These SFR limits can be directly compared to
those determined through optical measurements.  The average
dust-uncorrected {\lya}-estimated rate for the two galaxies is $\sim$7
{\sfr}, and that estimated from the UV continuum is slightly higher
$\sim$22 {\sfr} \citep{kod03}. While our average upper-limit of $\sim$
430 {\sfr} is an order of magnitude larger than this estimate, it is still a
useful first  constraint on the infrared properties of these systems; recall the results
of the submm and radio studies at $z\sim$ 3 that reveal star formation
rates in excess of $\sim$1000 M$_{\odot}$yr$^{-1}$ in systems for
which the UV prediction is $\sim$ 10 M$_{\odot}$yr$^{-1}$
\citep{cha05}.  The two LAEs discussed here are definitely not members
of such a luminous population.

\subsection{Dust Content}

Our deep submm imaging offers a unique opportunity to constrain the
mass of dust present in early starburst galaxies, a measurement that
has thus far only been possible at these redshifts for infrared luminous
quasars \citep{isa02,pri03}.  Using the same parameters we adopted in
\S3.1 (T = 36K, $\beta$ = 1.5), we calculate the dust mass following
\citet{hil83}:

\begin{equation}
M_d={S_{850{\mu}m} D_L^2\over(1+z)\kappa_\nu B_\nu (T)}
\end{equation}

\noindent where $\kappa_\nu$ is the dust emissivity, $\nu$ is the
rest-frame frequency, and B$_\nu$ is the Planck function at a
temperature T.  Following \citet{hil83}, we adopt $\kappa(125{\mu}m)$
= 1.875 m$^2$kg$^{-1}$ and scale with wavelength by $\kappa_\nu
\propto \nu^{\beta}$.  The results (see Table~\ref{submmtab}) are
consistent with the dust mass of the lensed Lyman-break galaxy MS
1512-cB58 which is roughly half the value of our upper limits for
similar dust parameters \citep{saw01}.  

Thus far, quasars are the only submm-detected systems that are
confirmed to lie at $z\sim$ 5-6 and  SED arguments suggest that their
submm luminosity is dominated by thermally emitting dust rather than
by non-thermal emission from the AGN \citep{pri03}.  The dust mass
estimates of the quasars range from a factor of 2 to 10 higher than
our $3\sigma$ upper limits (for the same dust parameters). Unless the dust temperature of the LAEs is much hotter than for AGN,  we would have detected objects with comparable dust masses in our data.
The lower dust content of the two LAEs is consistent with the
recent mid-infrared results of \citet{char05} who report evidence for
A$_\mathrm{v}$ = 1.0 mag of dust extinction in a single $z\sim$ 6.5
LAE, a value that is in the range of normal galaxies \citep{cho06}.

While a positive submm detection for either LAE would have provided dust formation
models with further empirical constraints, the non-detections discussed
here are unfortunately ambiguous. The existence of large dust masses in these LAEs would raise the
important question of how large amounts of dust can form in the
time between the formation of the first objects and $z\sim6.5$. 
Although the two $z\sim6.5$ LAEs have distinctly less dust
than the submm detected $z>$ 5 quasars, the $z\sim$ 2-3 SCUBA
population, and many of the $z\sim$ 2-3 Ly$\alpha$ blobs, we
cannot constrain  them to be less dusty than present-day normal
starburst galaxies.

The lack of a detection for either systems raises the following question: are
the $z\sim6.5$ LAEs even old enough to have built up a sizeable dust
reservoir through stellar evolution?  Perhaps large quantities of dust exists only in the oldest systems at $z>$ 6 and/or those
for which non-stellar dust production mechanisms are available: {\it i.e.}, quasars. 
 The Ly$\alpha$ equivalent-widths of the LAEs of 100-300\AA~(with
uncertainties of $>$ 50\%) imply ages of $\lesssim$ 10$^7$ years
\citep{mal02,cha93}, and this is  too young an age for large
amounts of dust to have been produced by  any known mechanism. Low mass stars have not had
sufficient time to evolve and  even
supernovae models have difficulty producing dust in such young objects
\citep{tod01}.  

Still, such large Ly$\alpha$ equivalent widths  have been measured for members of the submm
detected population at $z\sim$ 2-3 \citep{cha03}. These systems 
clearly have large dust masses but their Ly$\alpha$ EWs imply similar ages to the LAEs.  Thus while the Ly$\alpha$ EW may be a
reasonable indicator of the age of the current starburst, it cannot
provide complete information on the total star formation and dust
production history of a galaxy.  If the star formation histories of
the $z\sim$ 6.5 LAEs are extended over the first Gyr of the universe
 significant dust mass cannot be ruled out on the basis of the age of the 
 current starburst.

\subsection{Comparison to Lower Redshift Populations}

Using the submm upper-limit on the LAEs of
L$_\mathrm{FIR}<3\times10^{12}$L$_{\odot}$ results in an upper-limit
on the L$_\mathrm{FIR}$/L$_\mathrm{UV}$ ratio of $\lesssim35$
\citep{kod03}.  Our limit on the L$_\mathrm{FIR}$/L$_\mathrm{UV}$
ratio is substantially lower than the values of $\sim1000$ that are
observed for the submm-luminous population at $z\sim3$ \citep{cha05}.
Rather, a L$_\mathrm{FIR}$/L$_\mathrm{UV}$ ratio of $<35$ is within
the range of values exhibited by Lyman-break galaxies of
L$_\mathrm{FIR}$/L$_\mathrm{UV} \lesssim$ 10
\citep{ade00,web03a,gol02}; note that at $z\sim3$, LAEs make up
$\sim30$\% of the LBG population \citep{sha03}.

\begin{figure}
\plotone{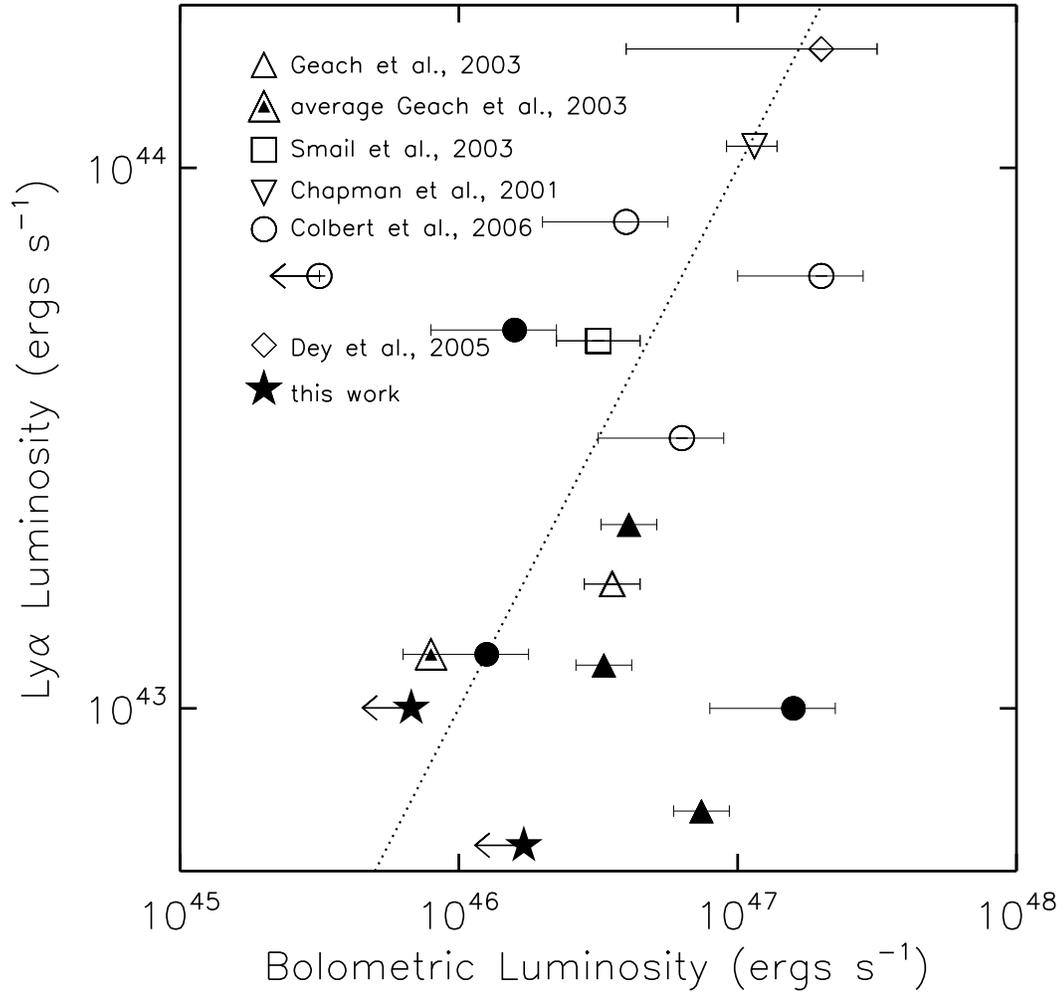}
\caption{The correlation between the flux in the Ly$\alpha$ line and
  the bolometric luminosity as inferred from single IR measurements at
  24$\mu$m or 850$\mu$m for extended Ly$\alpha$ emitters (LABs) at 2
  $<z<$ 3, compared to  the two SDF LAEs. For consistency   we have here adopted
  the same SED parameters as \citet{gea05} of T=40K, $\alpha$=4.5, $\beta$=1.7; however
 the parameters are similar to those discussed in the text and do not result in a significant change
 in L$_{\mathrm{bol}}$.     The dashed line is not a fit to the data but corresponds to
  the case where L(Ly$\alpha$) = 0.001 L$_\mathrm{bol}$ \citep{gea05}.
  Open points denote highly extended objects with sizes $>$ 50
  kpc$^2$, and solid points denote objects that are more compact
  than this.  \label{colbert}} 
\end{figure}

In contrast to the LAEs, many of the extended Ly$\alpha$ emitters at
lower redshifts, the so-called Ly$\alpha$ Blobs (LABs), are infrared
bright and have been detected by SCUBA or, more recently, at 24$\mu$m
by the Spitzer space telescope.  In Fig.~\ref{colbert}, we compare the Ly$\alpha$
luminosities to the bolometeric luminosities (assuming L$_\mathrm{IR}\sim$
L$_\mathrm{bol}$) of the two $z\sim$ 6.5 LAEs and   the lower redshift, infrared detected LAEs, as discussed in \citet{gea05} and \citet{col06}.  We divide the points into highly extended
(diameter$>$ 50 kpc) and compact (diameter$<$ 50 kpc) objects. There may be a weak tendency for the most extended  and Ly$\alpha$ luminous LABs to have
the highest infrared inferred bolometric luminosities, though note that two of the LABs with L$_{\mathrm{Ly}\alpha}$ $<$ 10$^{43}$ erg s$^{-1}$ have L$_\mathrm{bol}$ comparable to the most extended and luminous LABs.  The $z\sim6.5$ LAEs discussed here are
small ($\lesssim$ 4 kpc) with L$_\mathrm{Ly\alpha}$ 
comparable to the smaller IR bright LABs, but their L$_\mathrm{bol}$
are at least an order of magnitude lower.

A difference in the L$_\mathrm{bol}$ of different Ly$\alpha$ emitting systems, assumes however that all of the objects have the same dust temperature, and given the large differences in the extent of their Ly$\alpha$  emission this may not
be the case. If the spatial extent of the dust scales with that of the Ly$\alpha$ emission the temperature of the dust may also vary, such that the far-infrared flux of larger Ly$\alpha$ systems is dominated by cold diffuse dust, while the compact systems are dominated by similarly compact hot dust.   Given the same intrinsic source luminosity, 850$\micron$ measurements are biased toward cold dust and thus would not detect the hotter compact systems; that is, adopting a temperature which scales with source size could reduce or remove the trend  in Fig.\ref{colbert} for the 850{\micron} measurements. However, while this might explain the trend seen for the 850$\micron$ detected objects, such an effect cannot account for the properties of the 24$\micron$ detected systems, which appear to follow the same trend. For these objects L$_{\mathrm{bol}}$ is calculated from the rest-frame mid-infrared which does not suffer from the same bias as the 850$\mu$m data, and thus the smaller LAEs may truly be lower luminosity systems.  

An alternative explanation for the range in L$_\mathrm{bol}$ lies in the physics driving the intense energy production. 
Locally,  luminous and ultraluminous
infrared galaxies (U/LIRGs) are set apart from normal galaxies by their high merger
rate  and their tendency to host AGN, especially
at the high luminosity end (see \citet{san96} for a review). In these systems a major gas rich merger appears to be the trigger for both
star formation and AGN activity \citep[e.g.][]{mih96,bar96}.    The preponderance of AGN and/or merger signatures persist in infrared luminous systems at high-redshift such as the SMGs \citep{ale03,web03b,cons03} and some of the galaxies associated with LABs \citep[e.g.,][]{sma03,cha05,wei05,dey05,col06}. Infrared faint systems on the other hand, such as the SDF LAEs and the bulk of the LBG population \citep{web03a,hua05} do not, as a population, exhibit such properties \citep{ste96,hua05,leh05}.  This has lead many authors to conclude that the  violent infrared luminous phase of galaxy evolution is induced through gas rich mergers,  and thus systems such as the LAEs studied here are infrared-faint because they are not currently experiencing such an event.

\subsection{Star Formation Rate Density}

Using our SFR upper-limit and the luminosity function of LAEs at
$z\sim$ 6.5 presented by \citet{kas06}, we can estimate the maximum
contribution by these objects to the total SFRD at this redshift.  The
two LAEs discussed here have  $<\mathrm{L(Ly{\alpha}})>$ =
7.5$\times$10$^{42}$ ergs/s and $<\mathrm{SFR}> \lesssim$ 430
M$_{\odot}$yr$^{-1}$.  At this luminosity, they represent the brighter
end of the luminosity function as published by \citet{kas06} and have
a cumulative number density of $\sim$ 2$\times$10$^{-5}$ Mpc$^{-3}$. If we take
the SFR upper-limit determined here to be representative of the entire
LAE population above L(Ly${\alpha}$) = 2$\times$10$^{42}$ ergs/s,
the depth of the \citet{tan05} sample, we calculate their contribution
to the SFRD of to be $\lesssim$ 0.05 {\sfr} Mpc$^{-3}$.  We remind the reader
that this limit is based on a 3$\sigma$ upper-limit on the flux, and could
systematically shift by $\sim$70\% for a dust temperature range of $\pm$30\%.
 If we assume instead
that the SFR scales directly with L(Ly$\alpha$) and integrate over the
Ly$\alpha$ luminosity function of \citet{kas06} to L(Ly${\alpha})$ =
2$\times$10$^{42}$ ergs/s, this changes the upper-limit to $<$
0.03-0.3{\sfr} Mpc$^{-3}$, where the range corresponds to the range of
Schechter parameters describing the number density.

\begin{figure}
\plotone{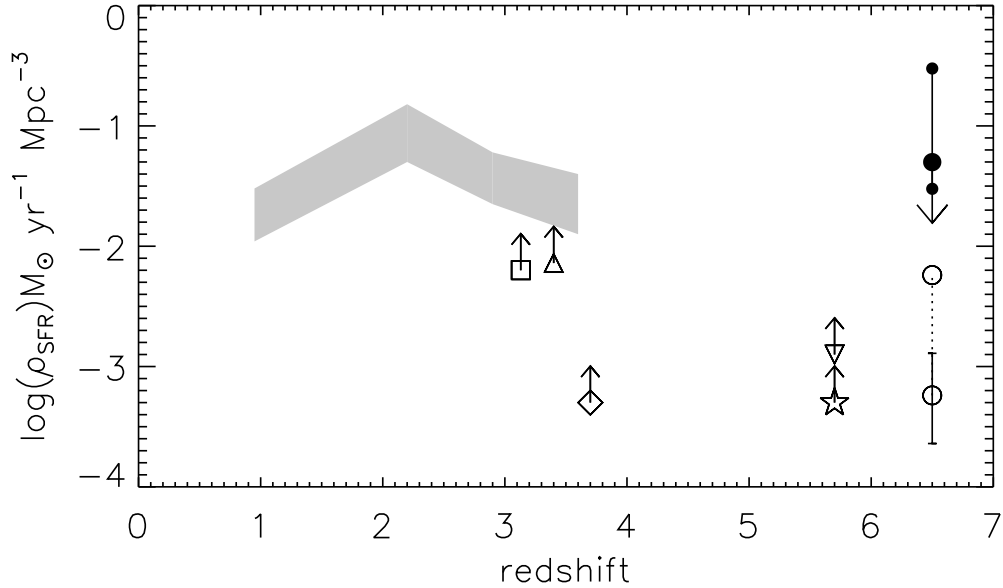}
\caption{The cosmic comoving star formation rate density (SFRD)  as a
  function of redshift. For clarity we show only two populations on
  this plot which are most relevant to this work:  the submm
  luminous galaxies (shaded area) and the LAEs (all other points). The
  lower bound on the submm shaded area corresponds to the measured
  SFRD from S$_{850{\mu}\mathrm{m}} > $ 5mJy galaxies and the upper
  bound shows the expected values when the population is  corrected
  for completeness down to  S$_{850{\mu}\mathrm{m}}  $ = 1mJy.  Points
  are \citet[][upside-down triange]{aji03},
  \citet[][star]{rho03},\citet[][triangle]{cow98}, \citet[][open
    square]{kud00}, and \citet[][diamond]{fuj03}. The lower open
  circle corresponds to the LAE population of \citet{tan05} from which
  the two LAEs studied here are drawn; and the upper open circle shows
  the revised SFRD of this population based on the dust extinction
  measured by \citet{char05}.  Our upper-limit, based on the
  \citet{tan05} number counts is shown by the large filled circle, and
  the range in upper-limits calculated from the Schecter-function fits
  of \citet{kas06} and assuming a linear relation between Ly$\alpha$
  line strength and IR-estimated SFR is shown by the two smaller solid
  points.  \label{sfr}} 
\end{figure}

In Fig.~\ref{sfr}, we compare our submm limits on the SFRD to
measurements from other LAE studies at $z>3$ as well as the submm
population at $z\sim2-3$ (see figure caption for references). The new
$z\sim$ 6.5 limits are comparable to that measured for the dust-obscurred
population at $z\sim2-3$ and thus does not constrain any turnover.  However, we stress that these are  upper
limits on the SFRD from only two LAEs at $z\sim6.5$.  Expanding our submm survey
to encompass the current sample of $z\sim6.5$ LAE candidates in the
SDF (58) and, $e.g.$ stacking the objects, would increase our depth by a factor of 5, and therefore begin to confirm a decline in the infrared-measured SFRD  from $z\sim3$ to $z\sim6$.

\section{Serendipitous Submm Sources}

Although no detection is made of the LAEs themselves, the submm maps
do contain a number of other sources (see Table~\ref{submmtab}).
SDF-1 in particular contains five objects detected above 3.5$\sigma$ at
850{\micron}, all of which are within 1 arcmin of the LAE.  This
represents a modest excess over the number counts expected from
confusion limited blank field surveys of $z \lesssim$ 2 sources
\citep[e.g.,][]{web03b}.  Whether the submm sources are at the same
redshift or possibly even at $z\sim6.5$ are intriguing possibilities.

At lower redshifts of $z\sim$ 3, there is tentative evidence from pair
counts and correlation analyses that the submm-luminous population is
strongly clustered; indeed, targeted imaging of rare, high-redshift systems such
as high-redshift radio galaxies \citep{ste03} have revealed
over-densities of submillimeter sources.  To explore the possibility
of whether the submm sources near the LAEs belong to a single massive
structure, either at the redshift of the two LAEs or in the
foreground, we attempt to constrain their redshifts.

\begin{figure}
\plotone{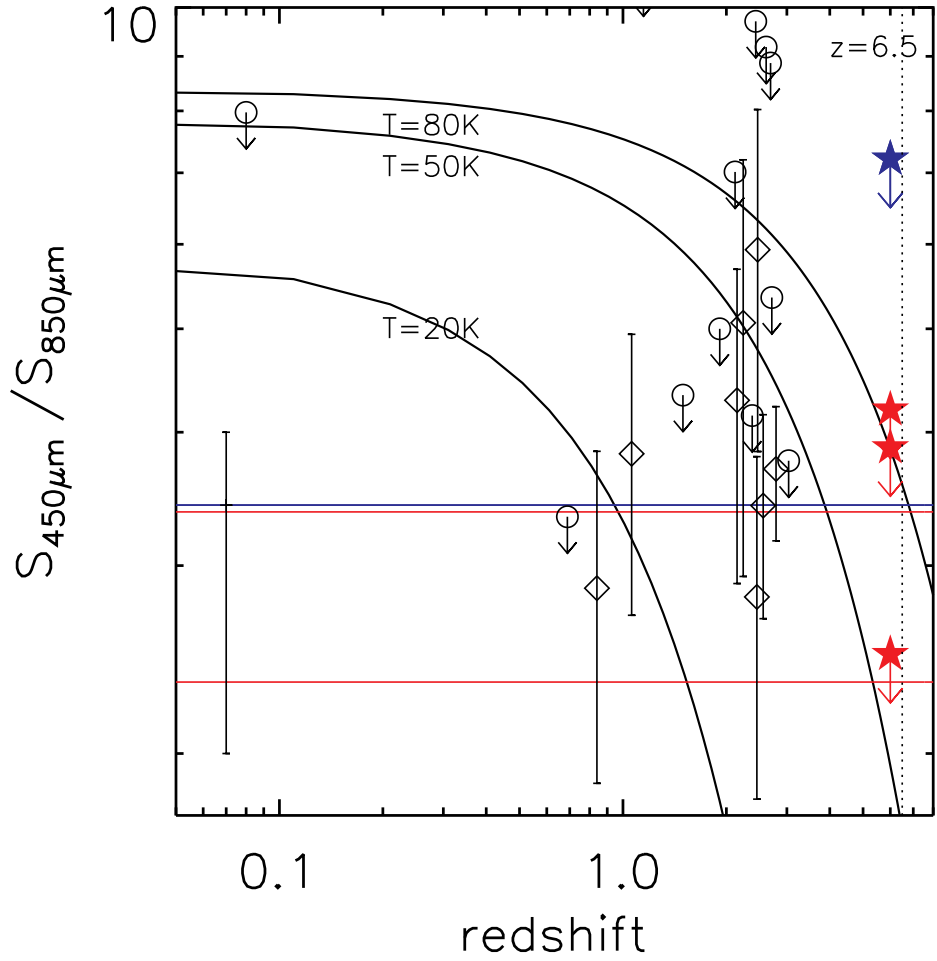}
\caption{The 450$\mu$m-850$\mu$m flux ratio as a function of
  redshift. The three curves correspond to three simple spectral
  energy distributions: grey body curves of varying dust temperature
  and an single dust emissivity of $\beta$=1.5.  The three horizontal
  lines show the ratio for the three 850$\micron$ and 450$\micron$
  detected sources, with the single error-bar in the lower left
  showing the uncertainty in the ratio (approximately equal for all
  three objects). The four solid stars denote 3$\sigma$ upper limits
  for the 850$\mu$m sources with no 450$\mu$m detection, arbitrarily
  placed at $z$ = 6 for clarity. Shown for comparison are the measured
  ratios (diamonds) or upper-limits (circles) for
  850$\micron$-selected sources with spectroscopic redshifts
  \citep{sco02,web03c,cle04,bor04,cha05}. \label{rat450}} 
\end{figure}

Two of the five 850{\micron} sources in the SDF-1 pointing and one in
the SDF-2 pointing are also detected at 450{\micron}, and their submm
flux ratios can be used as a very rough indication of their redshift
(Fig.~\ref{rat450}). Based on these ratios, it is unlikely that the
450$\mu$m-detected objects lie at the same redshift as the $z\sim6.5$
LAEs, unless they have extremely hot dust ($\sim$80K). Given a
reasonable temperature range, the submm sources lie between
1$<z<$5. The upper-limits on the four objects not detected at
450$\mu$m are consistent with this lower limit of $z>$ 1.

Considering the large uncertainties on the 450$\mu$m-850$\mu$m flux
ratios, it is possible that all of the submm sources lie at the same
foreground redshift, however, we cannot confirm nor deny this with
the current data-set. We have attempted to identify the optical
counterparts to these objects using the deep optical imaging available
on the SDF, but the number density of optical galaxies is simply too
great to allow unambiguous identifications.  Such an exercise requires
deep Spitzer imaging \citep{ash06} or radio observations
\citep{ivi02}.  We note that Spitzer imaging of the SDF will become
available in the next year, at which point we will attempt to identify
the optical/near-IR/mid-IR counterparts to the submm sources to better
constrain their redshifts.

\section{Conclusions} 

Quantifying the dust content in starburst galaxies at $z>6$ is
important both for determining cosmological properties such as the
star formation rate density and for testing stellar evolution models,
in particular how quickly large dust masses can form.  Using deep
submm observations of two spectroscopically confirmed Lyman-$\alpha$
emitting galaxies at $z\sim6.5$ in the SDF, we show that
these starburst galaxies do not contain large quantities of dust ($<$
4$\times$10$^8$ M$_\odot$).  Our S$_{850{\mu}m}$ imaging reaches the
 confusion limit and enables us to place  upper limits on
their star formation rates of $\lesssim$ 430 M$_\odot$ yr$^{-1}$.  The
two LAEs are markedly less infrared luminous and less dusty than the
submm detected population at $z\sim$ 2-3, the extended Ly$\alpha$
Blobs at $z\sim2-3$, and the $z>$ 5 submm detected quasars.  Rather,
the LAEs have L$_\mathrm{IR}$/L$_\mathrm{UV}$ ratios of $\lesssim$ 35
that are more in line with the Lyman-break population at lower
redshifts.

The submm observations at $z\lesssim3$ provide a  measurement 
of the star formation rate density at $z\sim$ 6.5 that is independent of the UV determined value.  Assuming the two LAEs are representative of the
entire $z\sim$ 6.5 LAE population and using the Ly$\alpha$ luminosity
function from \citet{kas06}, we can place an upper limit on the LAE
contribution to the dust enshrouded SFRD at $z=$ 6.5 of $<$ 0.05
M$_\odot$ yr$^{-1}$ Mpc$^{-3}$.  Although consistent with submm
measures of the SFRD at lower redshifts, and therefore unable to constrain a turnover,  we stress that this value is
extrapolated from only two objects.  More observations are clearly
needed. 

We recognize that because our results are based on only two $z\sim6.5$
LAEs, they are tentative.  However, our observations have yielded
the first  constraints on the dust content and total SF rates of such 
systems, and enable us to begin comparing the $z\sim6.5$ LAE galaxies
to the growing number of $z>5$ objects to better understand how these
early populations overlap.  Our deep submm imaging has also revealed a
 large number of serendipitously detected faint submm
sources that are likely to be at $1<z<5$.  A follow-up study on these
objects can provide much needed information on the faint ($S_{850{\mu}m}<$ 8mJy) submm
population.

Further constraints on dust enshrouded star formation at very high
redshifts are necessary and can be obtained through infrared
observations of larger samples of high-redshift LAEs. Indeed, our
program was originally intended to target a larger number of
$z\sim6.5$ LAEs, but it was cut short due to SCUBA's unfortunate
demise.  With the commissioning of SCUBA-2, such a study will again
become possible and will provide a more statistically representative
sample. In parallel, constraining the counterparts of
submillimeter-selected galaxies that do not have radio counterparts in
deep radio data will quantify the extent of the high-redshift ($z>$ 3) tail of
the submillimeter population.  Direct detections of dusty star
formation at high-redshift, to the level of a few 10{\sfr} (as are
currently predicted in the optical), also will be possible in the
future with the Atacama Large Millimeter Array.

\acknowledgments

We are grateful to L. Snijders and the JCMT staff for carrying out
part of the observations under the flexible observing program.  We
also thank K. Motohara and the Subaru Deep Team for the generous use
of their data.  Research by Tracy Webb acknowledges support from the NWO VENI
Fellowship program.  K. Tran and S. Lilly acknowledge support from the
Swiss National Science fund; K. Tran also acknowledges support from
the NSF Astronomy \& Astrophysics Postdoctoral Fellowship under award
AST-0502156. Finally, we thank the anonymous referee  whose careful reading of the 
manuscript and constructive comments greatly improved the paper.

\clearpage

\begin{deluxetable}{cccccccc}
\setlength{\tabcolsep}{0.025in}
\tabletypesize{\scriptsize}
\tablecaption{Submillimeter measurements\tablenotemark{a} of {\lya}-emitters
  \label{laetab}} 
\tablewidth{0pt}
\tablehead{
\colhead{Name}  & \colhead{$z$} & \colhead{S$_{850{\mu}\mathrm{m}}$(mJy)} & 
\colhead{S$_{850{\mu}\mathrm{m}}$(mJy)} & 
\colhead{S$_{450{\mu}\mathrm{m}}$ (mJy)}  &
\colhead{SFR$_{\mathrm{IR}}$ (M$_\odot$yr$^{-1}$)\tablenotemark{b}} &
\colhead{SFR$_{\mathrm{Ly\alpha}}$(M$_\odot$yr$^{-1}$)\tablenotemark{c}}
& \colhead{M$_{\mathrm{d}}$ 10$^8$M$_{\odot}$}  \\
\colhead{} & \colhead{} & \colhead{measured} & \colhead{limit} & 
\colhead{limit} & \colhead{} & \colhead{} & \colhead{}
}
\startdata 
SDF J132415.7+273058 (SDF-1) & 6.541 & -0.027 & $<$ 1.5 & $<$ 9.0 & $<$ 248 & 9.1 $\pm$ 0.8 & $<$ 2.3 \\
SDF J132418.3+271455 (SDF-2) & 6.578 & -1.0 & $<$ 3.8 & $<$ 9.5 & $<$ 613 &  5.1 $\pm$ 0.2 & $<$ 5.7 \\
\enddata
\tablenotetext{a}{3$\sigma$ upper limits are given; see \S 2.2.1 for
  discussion of the method.}  
\tablenotetext{b}{See \S 3.1 for description of the star formation rate
  estimate.} 
\tablenotetext{c}{Values taken from \citet{kod03}}
\end{deluxetable}


\begin{deluxetable}{cccccccc}
\tabletypesize{\scriptsize}
\tablecaption{Source Catalogs for the two LAE fields
  \label{submmtab}} 
\tablewidth{0pt}
\tablehead{
\colhead{Name}  & \colhead{RA (J2000)} &
\colhead{Dec (J2000)} & 
\colhead{S$_{850{\mu}\mathrm{m}}$ (mJy)}  & \colhead{S/N-850$\mu$m} &
\colhead{S$_{450{\mu}\mathrm{m}}$ (mJy)} & \colhead{S/N-450$\mu$m} & 
\colhead{$\Delta$ position} \\
\colhead{} & \colhead{} & \colhead{} & \colhead{} & \colhead{} & 
\colhead{} & \colhead{} &\colhead{850$\mu$m-450$\mu$m ($\arcsec$)}
}
\startdata
SDF1-850-1\tablenotemark{a} & 13:24:17.80 & 27:30:41.5 & 5.1 & 7.4 & 11.9 & 2.8 & 5.3 \\
SDF1-850-2 & 13:24:14.08 &  27:31:06.5 & 3.8 & 6.5 &  12.8 & 3.5 &4.0 \\
SDF1-850-3 & 13:24:17.77 & 27:31:33.5 & 4.6 & 6.3 & $<$ 11.4 & ... & ... \\
SDF1-850-4 & 13:24:15.21 & 27:31:34.0 & 3.1 & 4.8 & $<$ 12.0  & ... & ... \\
SDF1-850-5 & 13:24:16.38 & 27:30:00.0 & 3.0 & 4.0 & $<$ 12.6 & ... & ...\\
SDF1-450-1 & 13:24:18.33 & 27:30:34.5 & $<$ 2.0 & ... & 15.1 & 4.1 & 10.0 \\ 
SDF1-450-2 & 13:24:18.25 & 27:30:16.0 & $<$ 2.2 & ... & 16.8 & 4.3 &\\
SDF2-850-1 & 13:24:20.46 & 27:15:28.8 & 5.5 & 5.5 & 18.8 & 4.4 & 4.0 \\
SDF2-850-2 & 13:24:13.44 & 27:15:09.8 & 4.1 & 3.9 & $<$ 30.3 & ... & ... \\
SDF2-450-1 & 13:24:15.63 & 27:14:42.2 & $<$ 2.8 & ... & 16.0 & 3.5 & ...\\
\enddata
\tablenotetext{a}{There are two possible 450$\mu$m counterparts to
  this source. The first, which is listed here, is a 2.8$\sigma$ peak
  at an offset of $\sim$5{\arcsec} from the 850{\micron} position. The
  second possibility is source SDF1-450-1 (also listed) which is a
  $>$4.0$\sigma$ detection but at a 10$\arcsec$ offset. Please see the
  text \S 2.2 for discussion.}   
\end{deluxetable}

\clearpage

\end{document}